\documentclass[%
 reprint,
 superscriptaddress,
 nofootinbib,
 amsmath,
 amssymb,
 aps,
]{revtex4-2}

\usepackage{graphicx}% Include figure files
\usepackage{dcolumn}% Align table columns on decimal point
\usepackage{bm}% bold math
\usepackage{hyperref}% add hypertext capabilities
\usepackage[all]{hypcap} 
\usepackage[usenames,dvipsnames]{xcolor}
\hypersetup{colorlinks=true,
	citecolor=Blue,
	linkcolor=Magenta,
    	filecolor=NavyBlue,
	urlcolor=NavyBlue}

\def\be{\begin{equation}}
\def\ee{\end{equation}}
\def\ba{\begin{eqnarray}}
\def\ea{\end{eqnarray}}

\newcommand{\D}{\nabla}
\def\d{\mathrm{d}}

\def\({\left(}
\def\){\right)}
\def\[{\left[}
\def\]{\right]}
\def\L{\mathcal{L}}

\def\nn{\nonumber}

\begin{document}

\title{Probing massive fields with multiband gravitational-wave observations}

\author{Mu-Chun Chen}
\email{chenmuchun23@mails.ucas.ac.cn}
\affiliation{International Centre for Theoretical Physics Asia-Pacific, University of Chinese Academy of Sciences, Beijing 100190, China}
\affiliation{School of Astronomy and Space Science, University of Chinese Academy of Sciences, Beijing 100049, China}

\author{Hao-Yang Liu}
\email{liuhaoyang19@mails.ucas.ac.cn}
\affiliation{International Centre for Theoretical Physics Asia-Pacific, University of Chinese Academy of Sciences, Beijing 100190, China}
\affiliation{School of Physics, University of Chinese Academy of Sciences, Beijing 100049, China}

\author{Qi-Yan Zhang}
\email{zhangqiyan22@mails.ucas.ac.cn}
\affiliation{International Centre for Theoretical Physics Asia-Pacific, University of Chinese Academy of Sciences, Beijing 100190, China}

\author{Jun Zhang}
\email{zhangjun@ucas.ac.cn}
\affiliation{International Centre for Theoretical Physics Asia-Pacific, University of Chinese Academy of Sciences, Beijing 100190, China}
\affiliation{Taiji Laboratory for Gravitational Wave Universe (Beijing/Hangzhou), University of Chinese Academy of Sciences, Beijing 100049, China}

\begin{abstract}
We investigate the prospect of probing massive fields and testing gravitational theories with multiband observations of gravitational waves emitted from coalescing compact binaries. Focusing on the dipole radiation induced by a massive field, we show that multiband observations can probe the field with mass ranging from $10^{-16} $ to $10^{-15} {\rm eV}$, a parameter space that cannot be probed by the millihertz band observations alone. Multiband observations can also improve the constraints obtained with the LIGO-Virgo-KAGRA binaries by at most 3 orders of magnitude in the mass range. Moreover, we show that multiband observations can discriminate the spin of the field, which cannot be identified with single-band observations.
\end{abstract}

\maketitle

\section{Introduction}
\label{sec:intro}

The existence of particles beyond the standard model is not only predicted by fundamental theories, but also implied by observations of dark matter and dark energy. These new particles may weakly couple to the standard model particles, but could still get excited in extreme gravitational environments. It is known that fields of the new particles, such as light axions, generalized Proca, and extra degrees of freedom in gravity beyond general relativity (GR), can be significantly excited by neutron stars and black holes due to instability~\cite{Dima:2020yac,Herdeiro:2020wei,Dolan:2007mj,Percival:2020skc,Garcia-Saenz:2021uyv}, nonlinearity~\cite{Damour:1993hw,Doneva:2021tvn}, and other mechanisms~\cite{Sotiriou:2013qea,Silva:2017uqg, Hook:2017psm}, providing promising ways of searching for physics beyond the standard model. For example, superradiantly excited axions around supermassive black holes can induce additional polarization of light, and hence are constrained with the Event Horizon Telescope~\cite{Chen:2019fsq, Chen:2021lvo}. 

In recent years, gravitational waves (GWs) emitted by coalescing compact binaries have became a valuable probe to the physics in extreme gravitational environments. Deviations from GR and the standard model are searched and constrained by analyzing GW signals observed by the LIGO-Virgo-KAGRA (LVK) Collaboration~\cite{LIGOScientific:2016lio,LIGOScientific:2018dkp,LIGOScientific:2019fpa,LIGOScientific:2020tif,LIGOScientific:2021sio} (also see, e.g., Refs.~\cite{Yunes:2016jcc, Wang:2021gqm, Niu:2022yhr, Zhu:2023wci,Lin:2024pkr} and the references therein). With such a success in territorial detections, space-borne GW interferometers, such as LISA, TianQin, and Taiji, are planned to launch by the mid-2030s. These space-borne detectors target GWs of millihertz frequencies, which are complementary to the LVK band, and hence will probe fundamental physics from different approaches.

Interestingly, about $10$\text{--}$1000$ stellar mass black hole binaries are expected to be observed in both the millihertz band and the LVK band~\cite{Sesana:2016ljz}, opening the prospect for multiband GW astronomy. These binaries first inspiral in the millihertz band for several years, and then reappear in the LVK band typically a few weeks before they merge. Because of the long persistence of the signal in the millihertz band, space-borne detectors can measure the masses and sky location of the binaries with great accuracy, while LVK and future territorial detectors can measure the GW amplitude better due to the high signal-to-noise ratio (SNR). Therefore, multiband detection shall significantly improve parameter estimation in GW sources, and will be ideal for probing fundamental physics~\cite{Vitale:2016rfr}, such as constraining post-Newtonian (PN) and post-Einsteinian (PE) deviations~\cite{Yunes:2009ke, Carson:2019rda, Carson:2019kkh, Gupta:2020lxa,Datta:2020vcj}, searching for dipole GW radiation~\cite{Barausse:2016eii, Chamberlain:2017fjl, Liu:2020nwz,Zhao:2021bjw}, measuring GW dispersion relation~\cite{Harry:2022zey, Baker:2022eiz}, performing consistency tests~\cite{Vitale:2016rfr, Gnocchi:2019jzp, Carson:2019kkh} and bounding alternative gravity theories~\cite{Gnocchi:2019jzp}.

In this work, we emphasize the prospect of probing new massive bosonic fields with multiband GW observations. For massless (or ultralight) bosonic fields, their effects on orbital dynamics can be captured by parametrized PN and PE formalisms and have been constrained with the Hulse-Taylor pulsar~\cite{Anderson:2019eay, KumarPoddar:2019jxe, Seymour:2020yle,Poddar:2023pfj}, LVK binaries~\cite{LIGOScientific:2021sio} and pulsar timing arrays~\cite{Zhang:2023lzt}. They can also be probed by observations of GWs from extreme mass-ratio inspirals in the millihertz band~\cite{Maselli:2021men}.

Detectability of such fields with future GW detectors and multiband observations is also forecasted in the literature~\cite{Toubiana:2020vtf, Maselli:2020zgv, Liu:2020moh,Gao:2022hsn}. Effects of a massive field, however, do not generally fit with the parametrized PN or PE formalism and should be treated separately. In particular, for fields with masses heavier than $10^{-16} \, {\rm eV}$, their effects are suppressed for inspirals in the millihertz band~\cite{Sagunski:2017nzb,Huang:2018pbu, Kuntz:2019zef, Liu:2020moh, Barsanti:2022vvl, Diedrichs:2023foj}. Nevertheless, they can be constrained with LVK binaries to certain accuracy, if their mass is below $10^{-11} \, {\rm eV}$~\cite{Zhang:2021mks}. In this work, we shall demonstrate that constraints on massive fields can be significantly improved by 3 orders of magnitude with multiband observations, especially for the fields with mass ranging in $10^{-16} \text{--} 10^{-15} \, {\rm eV}$, which cannot be well probed by space-borne GW detectors. Moreover, we find that multiband GW observations can distinguish the spin of the field, if its mass is within $10^{-16} \text{--} 10^{-15} \, {\rm eV}$, which is hardly done with single-band observations. We work in the units of $G=\hbar=c=1$.

\section{Inspirals with massive fields}

Though the excitation mechanism depends on the theories, the new field, once excited by the compact object, typically affects binary inspirals by mediating additional force and emitting additional radiation. While the force could modify the inspiral waveform at $0$ PN order, the additional radiation usually manifests at $-1$ PN order~\cite{Huang:2018pbu}, and hence is the main signature that we are after.

To be concrete, we start from a massive spin-0 field with following action:
\be\label{s0}
S=\int \d^4 x \sqrt{-g} \( \frac12 \D_\mu \varphi \D^\mu \varphi- \frac12 \mu^2\varphi^2  + \L_{\rm int}  \) 
\ee
where $\varphi$ is the spin-0 field of mass $\mu$, and $\L_{\rm int}$ denotes the interaction between the field and the matter fields. 
During early inspiral, the two compact objects in a binary can be treated as two nonrelativistic pointlike particles. In this case, the leading order interaction term is
\be\label{s0int}
\L_{\rm int} \simeq \sum_{i=1,2} \sqrt{2}\, Q_i \delta \({\bf x} - {\bf x}_i(t) \) \varphi(t, {\bf x}) \, ,
\ee
where $Q_i$ and $ {\bf x}_i(t)$ represent the charge and position of the $i$th object respectively. We shall also consider a massive spin-1 field, the action of which is
\be\label{s1}
S=\int \d^4 x \sqrt{-g} \( -\frac14F_{\mu\nu}F^{\mu\nu} - \frac12 \mu^2 A_\mu A^\mu + \L_{\rm int}  \)\,, 
\ee
where $F_{\mu\nu} = \D_\mu A_\nu - \D_\nu A_\mu$. The spin-1 field couples to matter fields through their current $J^\mu$, i.e., $\L_{\rm int} \propto A_\mu J^\mu$. Again, treating the binary as two nonrelativistic pointlike particles, the coupling becomes
\be\label{s1int}
\L_{\rm int} \simeq \sum_{i=1,2} Q_i \delta \({\bf x} - {\bf x}_i(t) \) A^0(t, {\bf x}) \, .
\ee
In principle, one can consider massive spin-2 field which presents, for example, in bigravity. However, defining energy flux is subtle in theories with two dynamical metrics~\cite{Grant:2022koa}. It is also possible that the graviton itself has a nonzero but tiny mass, and GR should be replaced by a massive gravity theory. In this case, nonlinearity is expected to be important within the so-called Vainshtein radius, a scale that is typically much larger than the size of stellar mass binary. As a results, deviations from GR, including radiations of extra degrees of freedom in massive gravity, are expected to be suppressed for stellar mass binaries. For these reasons, we shall focus on and demonstrate the multiband detection strategy with only spin-0 and spin-1 fields.

Given the actions~\eqref{s0} and~\eqref{s1}, one can calculate the energy flux carrying by the radiations of the massive fields. For circular orbits, the energy flux of dipole radiation is given by
\be\label{flux}
P_{\rm MF} = \frac{2}{3} \Delta q^2 \(\frac{m_1 m_2}{m_1+m_2}\)^2 R^2 \Omega^4\, g\(\Omega, \mu\)
\ee
with
\ba
g\(\Omega, \mu\) = \left \{
\begin{aligned}
& \[1-\(\tfrac{\mu}{\Omega}\)^2\]^{3/2}\,  & \text{spin-0} \\
& \sqrt{1-\( \tfrac{\mu}{\Omega} \)^2}  \[1+ \frac12 \(\tfrac{\mu}{\Omega}\)^2\]  & \text{spin-1}
\end{aligned} \right. \,,
\ea
where $R$ is the orbital separation, $\Omega$ is the orbital frequency and $\Delta q \equiv \(Q_1/m_1\)- \(Q_2/m_2\)$ with $m_{1,2}$ being the mass of compact objects (see, e.g., Ref.~\cite{Krause:1994ar} for details). The models we consider here are generic, and hence the energy flux calculated by Eq.~\eqref{flux} applies to many theories: see Refs.~\cite{Alsing:2011er, Yagi:2012gp, Yagi:2015oca,  Zhang:2019iim, Niu:2019ywx,  Liu:2020moh, Niu:2021nic, Liu:2022qcx, Li:2023lqz, Bhattacharyya:2023kbh, Bhattacharyya:2024aeq} for example.

To incorporate the effects of the dipole radiation from the massive fields into inspiral waveform, we calculate the waveform in frequency domain by extending the TaylorF2 template. Specifically, the waveform template is given in the frequency domain,
\ba\label{TF2}
h(f) \simeq H(f) \exp \left[ i \Psi(f) \right]\, ,
\ea
where $\Psi(f) =  2\pi f t-\phi -\tfrac{\pi}{4}$ is the GW phase, and is calculated under the stationary phase approximation,
\ba
 t(f)  &=& t_c - \int_{f_c}^{f} \frac{1}{P} \left(\frac{dE}{df'}\right) \d f' \,, \\
 \phi(f)  &=& \phi_c - \int_{f_c}^{f} \frac{2\pi f' }{P} \left(\frac{dE}{df'}\right) \d f'\,.
\ea
Here $E$ and $P$ are binding energy and total radiation power of the binary system, while $t_c$ and $\phi_c$ are the time and phase at merger. In the presence of a massive field, the total radiation power can be written as $P=P_{\rm GR} + P_{\rm MF}$. Namely, it includes energy fluxes of both GWs in GR and radiation of the massive field. Assuming $|\Delta q| \ll 1$, the massive field induces an extra phase $\Psi_{\rm MF}$ in the waveform,   
\be
\Psi(f) \approx \Psi_{\rm GR}(f) + \Psi_{\rm MF}(f),
\ee
where $\Psi_{\rm GR}$ is the GW phase predicted by GR, and the explicit expression of $\Psi_{\rm MF}$ is given in the  Appendix.~\ref{App:phase}.

\begin{figure*}
    \centering
    \includegraphics[width=0.75\linewidth]{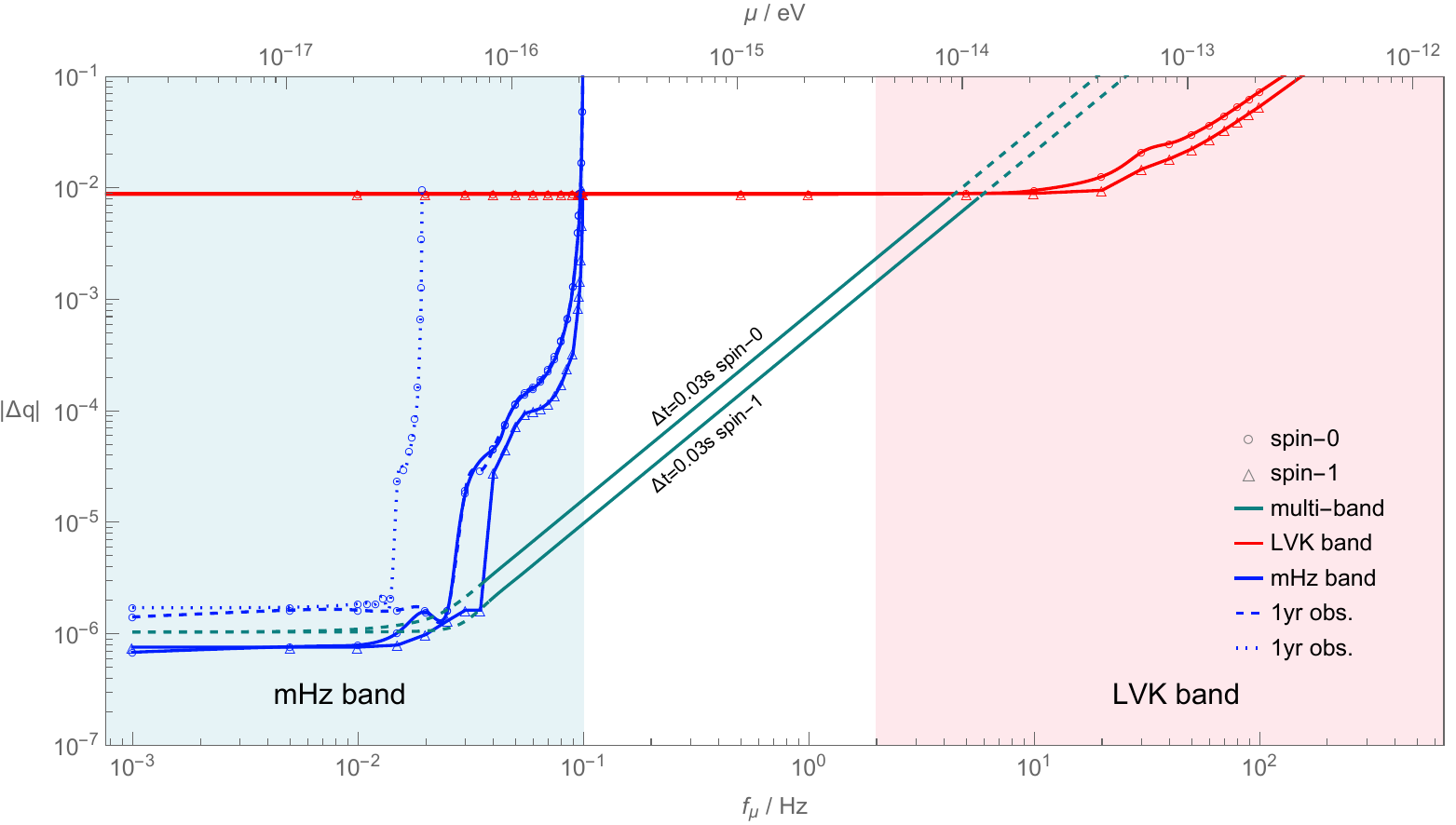}
     \caption{Detectability on spin-0 and spin-1 fields of different mass. The markers show the expectation errors on $\Delta q$ obtained by calculating the Fisher information matrix [cf. Eq.~\eqref{error}], assuming a GW150914-like event observed in the millihertz and the LVK bands. Here circles and triangles show the errors for spin-0 and spin-1 fields, respectively, while blue and red denote the millihertz and the LVK bands. For observations in the millihertz bands, markers jointed by the solid, dashed, and dotted lines show the situation of a $4$-yr observation starting at four years before merger, a $1$-yr observation starting at one year before merger, and a $1$-yr observation starting at four years before merger, respectively. The green lines show the merger time deviation from GR, when there is a spin-0 or a spin-1 massive field. }\label{fig:dq}
\end{figure*}

\section{Single-band observations}
\label{sec:single}

Given the inspiral waveform, we first investigate detectability of the massive fields with single-band observations. It is convenient to introduce
\ba
f_{\mu} \equiv \frac{\mu}{\pi} = 0.48 \times \left(\frac{\mu}{10^{-15} \, {\rm eV}}\right) {\rm Hz}\,.
\ea
In the LVK band, we expect to observe the binary coalescence up to merger, and the phase shift induced by dipole radiation of the massive field can be estimated as
\be\label{dPsi}
\Psi_{\rm MF} (f) \sim \frac{5}{2688} \Delta q^2 \nu^{-1} \( \pi M f \)^{-7/3}\,,
\ee
with $M=m_1+m_2$ and $\nu = m_1 m_2/M^2$. Here $f$ should be $f_\mu$ or $f_i$, whichever is larger, with $f_i \sim 2\,{\rm Hz}$ being the GW frequency when the inspiral signal enters the LVK band. Assuming the dipole radiation can be detected if $\Psi_{\rm MF} \sim {\cal O}(1)$, we expect that detectors in the LVK band are sensitive to $\Delta q \gtrsim 0.008$ for fields with $f_\mu < f_i$, and the sensitivity gets worse as $f_\mu$ increases. We shall also estimate the detectability with observations in the millihertz band. Different from the LVK band, we may not observe a clear chirping in GW frequency during the observation time, if the binary is in its very early inspiral stage when the detector turns on. In particular, the time to merger given by GR is
\ba
\tau_{\rm GR} &\simeq& \frac{5}{256} M \nu^{-1} \(\pi M f_i \)^{-8/3}  \\
&\approx& 1.03\,{\rm yrs} \(\frac{\nu}{0.25}\)^{-1} \(\frac{M}{60 M_\odot}\)^{-5/3} \(\frac{f_i}{\rm 0.03\,Hz}\)^{-8/3} \nn .
\ea
Therefore, if the binary is emitting GWs with $f_i \gtrsim 0.03 \,{\rm Hz}$ when detector turns on, the signal will leave the millihertz band during the observation period, which is typically assumed to be $1$ or $4$ yr. The phase shift induced by the dipole radiation during the observation time can also be estimated with Eq.~\eqref{dPsi}, and we expect the minimally detectable $\Delta q$ is $\sim 6 \times 10^{-5}$ for $f_\mu < f_i$. On the other hand, for $f_i \ll 0.03 \,{\rm Hz}$, we do not expect to see $f$ changing significantly during the operating time, and the detector can only probe the dipole radiation of fields with $f_\mu \lesssim f_i < 0.03\,{\rm Hz}$. In this case, the additional phase shift is given by
\be\label{earlyPsi}
\Psi_{\rm MF} \sim \frac{2}{21} \Delta q^2 (\pi M f_i)^{1/3} M^{-1} T_{\rm obs} \, ,
\ee
where $T_{\rm obs}$ is operating time. The minimally detectable $|\Delta q|$ is $\sim 3 \times 10^{-5}$ assuming a $4$-yr observation. Comparing to the LVK band, observations in the millihertz band generally have better accuracy for fields with $f_\mu \lesssim 0.01 \,{\rm Hz}$. However, the accuracy gets worse quickly as $f_\mu$ increases. In fact, observations in the millihertz band almost cannot probe fields with $f_{\mu} > 0.1\,{\rm Hz}$, because the massive field almost plays no role during the entire observation.

A rigorous forecast on detectability can be made using Fisher information matrix. Having in mind an interferometric detector and working in frequency domain, the measured data $d$ can be expressed as a linear combination of the signal $s$ and the detector noise $n$, 
\be
d(f) = s(f, {\bm \theta}) + n(f).
\ee 
The signal is the detector response to GWs that are characterized by a set of parameters ${\bm \theta}$. Assuming the noise to be stationary and Gaussian with a single-sided spectral density $S_n \(f\)$, the likelihood is
\be
{\cal L}\(d | {\bm \theta} \) \propto \exp \[-\frac12 \langle d - s({\bm \theta}) | d - s({\bm \theta}) \rangle \] \, ,
\ee
where the inner product is defined by
\be
\langle a|b \rangle = 2 \int_{-\infty}^{+\infty} \frac{a(f)b^*(f)+a^*(f)b(f)}{S_n(f)} \, \d f \, .
\ee
For signals with large SNR, the detectability of the signal characterized by ${\bm \theta}_0$ can be forecasted with Fisher information matrix,
\be
\Gamma_{ij}  = \left. \left\langle  \frac{\partial s }{\partial \theta_i  } \right| \left. \frac{\partial s }{\partial \theta_j  }  \right\rangle \right|_{{\bm \theta} ={\bm \theta}_0 } \, .
\ee
The expectation value of the errors are given by
\be\label{error}
\sigma_{\theta_i} = \(\Gamma^{-1}\)_{ii}\,.
\ee

For demonstration, we consider a GW150914-like event, i.e., a binary with $m_1=36 M_\odot$ and $m_2=29 M_\odot$ inspiralling at $410 \,{\rm Mpc}$, and estimate the detectability of $|\Delta q|$ by calculating the Fisher information matrix for a spin-0 field with certain $\mu$. For simplicity, we average over the binary's sky location, inclination, and polarization, and ignore the spin effects for simplicity. Then the parameters reduce to
\be
{\bm \theta} = \{{\cal M}, \eta, d_{L}, t_c, \phi_c, \Delta q \}\, ,
\ee
which are the chirp mass, the dimensionless reduced mass, the luminosity distance, the merger time, the coalescence phase and the charge difference respectively. 

For observations in the LVK band, the Fisher information matrix is calculated with the updated advanced LIGO design sensitivity curve~\cite{ligocurve}. As shown in Fig.~\ref{fig:dq}, such an observation can probe $|\Delta q| > 10^{-2}$ for $f_\mu < 5\,{\rm Hz}$, and the sensitivity on $\Delta q$ becomes worse as $f_\mu$ approaches the chirp frequency, which is $\sim 200 \,{\rm Hz}$.

To forecast the detectability in the millihertz band, we use the effective sensitivity curve of LISA~\cite{Robson:2018ifk} obtained after averaging over sky and polarization angles. Given an GW150914-like event, we may have different situations, and the detectability on $\Delta q$ is shown in Fig.~\ref{fig:dq}. We first consider a $1$-yr observation, starting at $1$ yr before the binary merges. In this case, we have $f_{i} \simeq 0.029 \,{\rm Hz}$. With such an observation, LISA can detect $\Delta q > 2\times10^{-6}$ for $f_\mu < 0.029 \,{\rm Hz}$. We then consider a $4$-yr observation, starting at $4$ yr before the binary merges, in which case $f_{i} \simeq 0.017 \,{\rm Hz}$ and LISA can detect $\Delta q >  8\times10^{-7}$ for $f_\mu < 0.017 \,{\rm Hz}$. We find that, for $f_\mu \ll f_i$, the detectability approximately improves with the observation duration, which is expected from Eq.~\eqref{earlyPsi}. In both cases, the detectability becomes worse quickly as $f_\mu$ approaches $0.1 \,{\rm Hz}$, as beyond which the field barely affects the orbital dynamics in the observation period. In particular, detectability of $1$-and $4$-yr observations is almost the same for $f_\mu > 0.02 \,{\rm Hz}$, because the massive field only becomes dynamically relevant in the last year of the observation. For comparison, we also consider a $1$-yr observation, but starting at $4$ yr before the binary merges. Given such an observation, LISA can detect $\Delta q > 2\times10^{-6}$ for $f_\mu < 0.01 \,{\rm Hz}$, and quickly loses the detectability as $f_\mu$ approaches to $ 0.017 \,{\rm Hz}$, as beyond which the mass field becomes irrelevant to inspiral during the observation period. We perform a similar analysis for a spin-1 field as well and show the results in Fig.~\ref{fig:dq}.

\section{multiband observations}

\begin{figure}[t]
    \centering
    \includegraphics[width=0.9\linewidth]{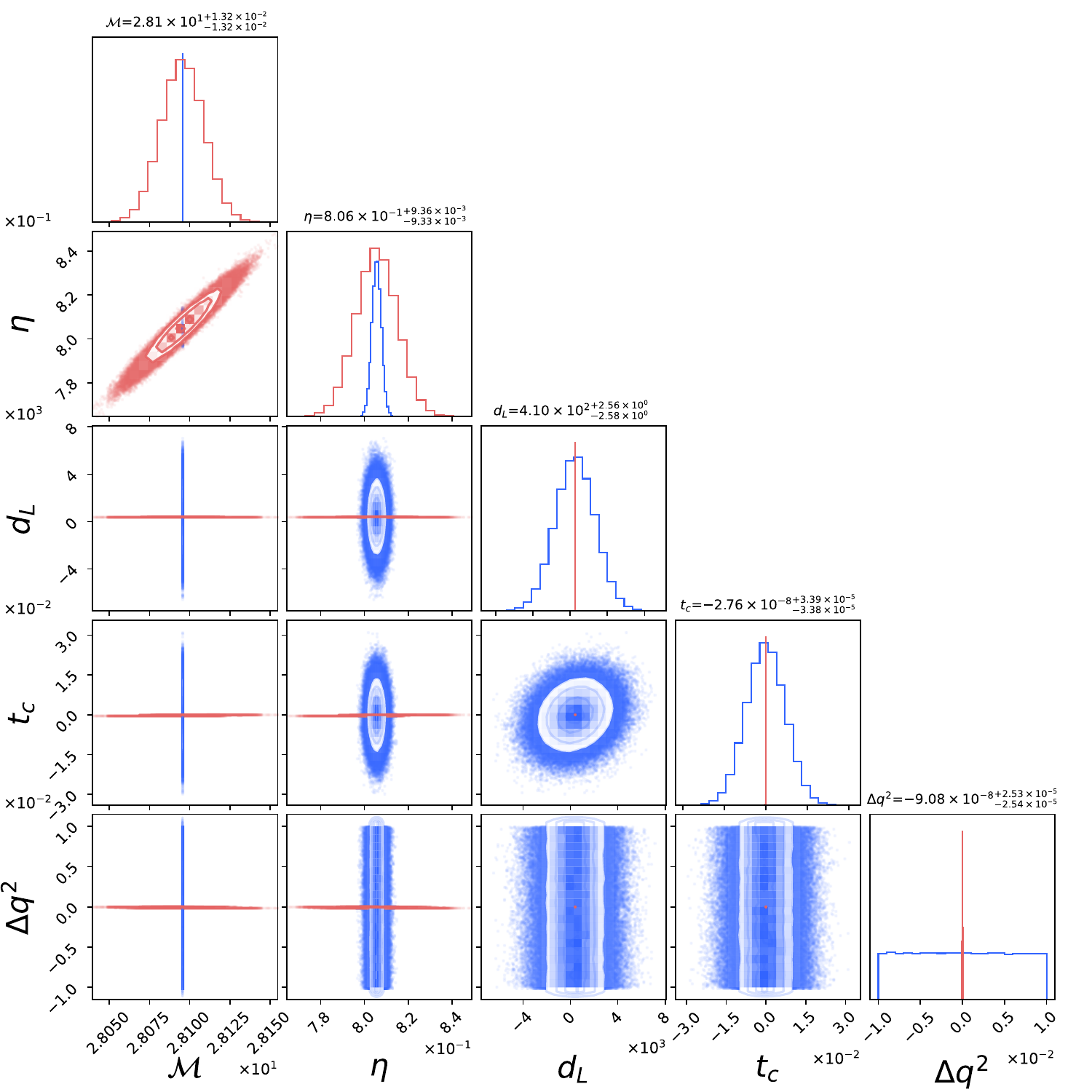}
     \caption{Parameter constraints for a GW150914-like event, i.e., $m_1=36 M_\odot$, $m_2=29 M_\odot$ and $d_L=410 {\rm Mpc}$,  obtained with millihertz band (blue) and LVK band (red) observations. When calculating the Fisher information matrix, we assume the presence of a spin-0 field with $f_\mu = 1 \,{\rm Hz}$.}\label{fig:corner}
\end{figure}

\begin{figure*}
    \centering
    \includegraphics[width=0.75\linewidth]{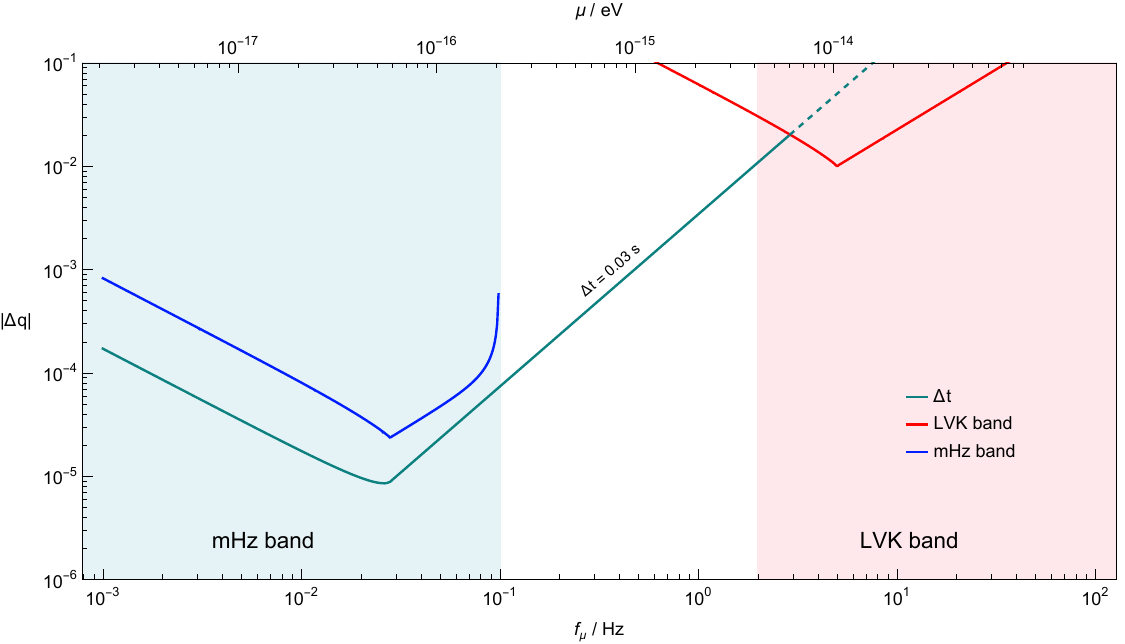}
     \caption{Discrimination on the spin of massive fields. The blue (red) lines show the difference between the spin-0 and the spin-1 fields in GW phase, $\Delta \Psi \equiv \left| \Psi_{\rm MF}^{s=0} - \Psi_{\rm MF}^{s=1} \right|$, assuming a GW150914-like event observed in the millihertz (LVK) band. The green line shows the difference between the spin-0 and the spin-1 field in merger time $\Delta t$. Region above the green line and below the blue and red lines is the parameter space where the detectability on the spin of the field is expected to be improved by multiband observations.}\label{fig:ds}
\end{figure*}

As demonstrated previously, observations in the millihertz band alone can improve the detectability on massive fields, but only for fields with $f_{\mu} < 0.1 \,{\rm Hz}$. Nevertheless, fields with $f_\mu > 0.1 \,{\rm Hz}$ can still affect the later binary evolution and accelerate orbital decay while observations in the millihertz band can forecast the merger time with an accuracy of ${\cal O}(1)$ s. With a multiband observation that the binary is observed later in the LVK band, the merger time forecasted by the millihertz observations can be tested in the LVK band, providing an additional constraint on the dipole radiation induced by the massive fields. In Fig.~\ref{fig:corner}, we consider a GW150914-like event and estimate the constraints of the waveform parameters assuming there is a spin-0 field with $f_\mu = 1 \,{\rm Hz}$. The constraints are obtained using the Fisher information matrix as stated in Sec.~\ref{sec:single}. Given the mass of the field, LVK band observation can probe the field if $|\Delta q| > 10^{-2}$, while the millihertz band observation alone cannot probe the field at all. For other parameters, observation in the LVK band does better measurement on distance and merger time, while observation in the millihertz band does better measurement on masses. In particular, the millihertz band can measure the merger time with an accuracy of $0.03 $\, s at $90\%$ confidence level. 

Assuming $|\Delta q| \ll 1$, the merger time in the presence of dipole radiation will be earlier than that in GR by an amount of 
\ba
\Delta t &\simeq& \int_{f_\mu}^{f_c} \d f \(\frac{P_{\rm MF}}{P_{\rm GR}^2}\) \(\frac{\d E}{\d f}\) \\
&\sim& \frac{5}{3072} \Delta q^2 \nu^{-1} \(\pi M f_\mu\)^{-10/3}\, .
\ea
By requiring $\Delta t < 1 \,{\rm s}$, we find a constraint of
\be
|\Delta q| < 10^{-4} \left(\frac{\nu}{0.25}\right)^{1/2} \left(\frac{M}{60 M_\odot}\right)^{5/3} \left(\frac{f_\mu}{1\,{\rm Hz}}\right)^{5/3}\,,
\ee
for $ 0.1 < f_{\mu} <  2 \,{\rm Hz}$.
In Fig.~\ref{fig:dq}, we show the constraints on $|\Delta q|$ for fields with different masses by requiring $\Delta t < 0.03 \,{\rm s}$, assuming a multiband observation of a GW150914-like event. Comparing to the single-band observations, we find that the multiband observation indeed improves the detectability on $\Delta q$ by filling the gap between the millihertz and the LVK bands.

In addition to improving detectability of the massive fields, multiband observation can further distinguish spin-0 fields from spin-1 fields. With observations in the LVK band or in the millihertz band alone, we expect to distinguish the fields if they could induce ${\cal O}(1)$ difference in GW phase, namely, $\left| \Psi_{\rm MF}^{s=0} - \Psi_{\rm MF}^{s=1} \right| \sim {\cal O}(1)$. Given Eq.~\eqref{flux}, waveforms of spin-0 and spin-1 fields differ notably only when the dipole radiation just turns on. Therefore, if $\mu \ll \Omega$ during the single-band observation, one cannot distinguish whether the dipole radiation is caused by a spin-0 or spin-1 field, cf. Eq.~\eqref{flux}, even with a nontrivial measurement of $\Delta q$. With multiband observations, one can further infer whether the dipole radiation observed in the LVK band is from a spin-0 or spin-1 field by examining the merger time. Considering the GW150914-like event, Fig.~\ref{fig:ds} shows the GW phase difference for observations in the LVK and the millihertz band, as well as the difference in merger time $\Delta t=|t_c^{s=0}-t_c^{s=1}|$, from which we can conclude that multiband observations can distinguish the spin of the fields in some of the parameter space where the single-band observation cannot.

\section{Conclusion and Discussion}

In this work, we investigate the prospect of probing massive fields with space-borne detectors that target GWs in the millihertz band. We consider the dipole radiation from generic massive spin-0 and spin-1 fields, calculate their imprints on the inspiral waveform and estimate the detectability on the fields. We demonstrate the detectability on the effective charge difference $|\Delta q|$ of LISA by calculating the Fisher information matrix. We find that LISA can constrain $|\Delta q|$ down to $10^{-6}$ for fields with mass below $3 \times 10^{-17} \,{\rm eV}$, given a 4-yr observation of a GW150914-like binary. Note that we use GW150914-like event as a benchmark to demonstrate the sensitivity of $|\Delta q|$, while the sensitivity does depend on the binary parameters. For fields with $f_{\mu} > 1 {\rm Hz}$, the sensitivity is mainly determined by the number of orbital cycles observed in the LVK band, and hence will get improved if the binary has a smaller total mass. Constraints on fields with $10^{-1} {\rm Hz} < f_\mu < 1 {\rm Hz}$ replies on the precision of merger time measured in the millihertz band, which distributes from $0.1$s to $100$s and peaks around $3$s~\cite{Sesana:2017vsj}. As the precision of merger time mostly depends on the signal-to-noise ratio, the sensitivity of $|\Delta q|$ can be improved if the event has a larger signal-to-noise ratio in the millihertz band. For fields with $f_{\mu} < 10^{-1} \,{\rm Hz}$, the sensitivity is mainly determined by the number of orbital cycles observed in the millihertz band. Given a certain observation time, the sensitivity will be improved if the binary has smaller total mass. In practice, the massive field will be probed by stacking multiple events. The sensitivity of $|\Delta q|$ given multiple multiband observations will be investigated in future work. In addition, we assume circular orbits in our analysis, which is usually the case in the LVK band, but is not necessarily true in the millihertz band. Depending on the formation channel, stellar mass binaries observed in the millihertz band could have nonzero eccentricity, for example, see Refs.~\cite{Cardoso:2020iji, Zhao:2023ilw} and the references therein. Since both scalar and vector radiations tend to circularize binaries~\cite{Cardoso:2020iji}, the radiation power and hence $|\Delta q|$ could be probed by measuring the eccentricity distribution, if the field mass is relatively small such that the radiation turns on in the millihertz band. For fields with larger mass, one cannot probe the fields with eccentricity, which is washed out by GW radiation anyway.

As mentioned in Sec.~\ref{sec:intro}, a direct coupling to a spin-0/spin-1 field is constrained by other experiments. For example, precision measurements of binary pulsars impose a constraint of $|\Delta q| < 10^{-3}$ for fields with mass lower than $10^{-16} \, {\rm eV}$~\cite{Seymour:2020yle}. Also, the direct coupling can, in principle, lead to a fifth force, which is constrained by local gravity tests~\cite{Tsujikawa:2008uc}. If we assume the charge difference is of the same order of the charges, the Eöt-Wash experiment indicates a constraint of $|\Delta q| < 10^{-1}$ for fields with mass lower than $10^{-3} \,{\rm eV}$ (see, for example, Ref.~\cite{Kapner:2006si} for the accurate plot of the constraints). One should note that the coupling strength is highly model dependent. It can vary in different systems even assuming the same theory. Therefore, constraints obtained in one experiment does not directly apply to the others.

We further emphasis the implication of multiband observations on probing new massive fields. We show that multiband observation of a GW150914-like binary can improve the detectability on $\Delta q$ by at most 3 orders of magnitude for spin-0 and spin-1 fields with mass ranging from $10^{-16}$ to $10^{-15}$\,${\rm eV}$. Multiband observation can further distinguish the spin of the fields in such mass range, where the spin of the field cannot be identified with single-band observations even if a nontrivial $\Delta q$ is detected.

Our multiband detection strategy mainly depends on the measurement of merger time, which could, in principle, be affected by other beyond GR effects, such as modification of GW propagation speed. We expect such degeneracy can be broken by considering multiple multiband events, because effects from modification of GW speed depends on the distance of the sources while effects from massive fields should depend on the intrinsic parameters of the merger binaries. We shall leave this topic for future investigation.

\begin{acknowledgments}
We thank Lijing Shao and Yuetong Zhao for helpful discussions. J. Z. is supported by the scientific research starting grants from University of Chinese Academy of Sciences (Grant No.~E4EQ6604X2), the Fundamental Research Funds for the Central Universities (Grants No.~E2EG6602X2 and No.~E2ET0209X2), and the National Natural Science Foundation of China (NSFC) under Grants No.~12147103 and No.~12347103.
\end{acknowledgments}

\appendix

\section{Corrections on GW phase}
\label{App:phase}

In this appendix, we show the explicit expressions for the corrections on the GW phase in the TaylorF2 waveform. As stated in the main text, the waveform template in frequency domain is
\ba
h(f) \simeq H(f) \exp \left[ i \Psi(f) \right]\, ,
\ea
where $\Psi(f) =  2\pi f t-\phi -\tfrac{\pi}{4}$ is the GW phase and is calculated under the stationary phase approximation,
\ba
 t(f)  &=& t_c - \int_{f_c}^{f} \frac{1}{P} \left(\frac{dE}{df'}\right) \d f' \,, \\
 \phi(f)  &=& \phi_c - \int_{f_c}^{f} \frac{2\pi f' }{P} \left(\frac{dE}{df'}\right) \d f'\,.
\ea
Here $E$ and $P$ are the binding energy and total radiation power of the binary system, while $t_c$ and $\phi_c$ are the time and phase at merger. In the presence of a massive field, the total radiation power $P=P_{\rm GR} + P_{\rm MF}$. Assuming $|\Delta q| \ll 1$, we have $P_{\rm MF} \ll P_{\rm GR}$, and hence $t(f) \approx t_{\rm{GR}}(f) + t_{\rm{MF}}(f)$ and $\phi(f) \approx \phi_{\rm{GR}}(f) + \phi_{\rm{MF}}(f)$ with
\begin{align}
t_{\rm{MF}}(f) &=\int\frac{P_{\rm{MF}}}{{P_{\mathrm{GR}}^2}}\frac{\mathrm{d}E}{\mathrm{d}\Omega}\mathrm{d}\Omega\,,\\
\phi_{\rm{MF}}(f) &=\int^{f}_{f_c}\frac{2\pi fP_{\rm{MF}}}{{P_{\mathrm{GR}}^2}}(\frac{\mathrm{d}E}{\mathrm{d}f})\mathrm{d}f\,.
\end{align}
For the spin-0 field, we have
\begin{align}
    t_{\rm{MF}}^{s=0}(f)&=\frac{3{f_\mu}^2}{19f^{\frac{16}{3}}_c}\sqrt{\frac{1-\frac{{f_\mu}^2}{f^2_c}}{1-\frac{f^2_c}{{f_\mu}^2}}}\times{}_2F_1\left(-\frac{19}{6}, -\frac{3}{2}; -\frac{13}{6}; \frac{f^2_c}{{f_\mu}^2}\right)\nonumber\\
    &-\frac{3{f_\mu}^2}{19f^{\frac{16}{3}}}\sqrt{\frac{1-\frac{{f_\mu}^2}{f^2}}{1-\frac{f^2}{{f_\mu}^2}}}\times{}_2F_1\left(-\frac{19}{6}, -\frac{3}{2}; -\frac{13}{6}; \frac{f^2}{{f_\mu}^2}\right)\,,\\
    \phi_{\rm{MF}}^{s=0}(f)&=\frac{3{f_\mu}^2}{16f^{\frac{13}{3}}_c}\sqrt{\frac{1-\frac{{f_\mu}^2}{f^2_c}}{1-\frac{f^2_c}{{f_\mu}^2}}}\times{}_2F_1\left(-\frac{8}{3}, -\frac{3}{2}; -\frac{5}{3}; \frac{f^2_c}{{f_\mu}^2}\right)\nonumber\\
    &-\frac{3{f_\mu}^2}{16f^{\frac{13}{3}}}\sqrt{\frac{1-\frac{{f_\mu}^2}{f^2}}{1-\frac{f^2}{{f_\mu}^2}}}\times{}_2F_1\left(-\frac{8}{3}, -\frac{3}{2}; -\frac{5}{3}; \frac{f^2}{{f_\mu}^2}\right)\,.
\end{align}
For the spin-1 field, we have
\begin{align}
    t_{\rm{MF}}^{s=1}(f) &= \frac{3}{494f^{16/3}_c} \sqrt{\frac{1 - \frac{{f_\mu}^2}{f^2_c}}{1 - \frac{f^2_c}{{f_\mu}^2}}} \times \left[13 \sqrt{1 - \frac{f^2_c}{{f_\mu}^2}} (f^2_c - {f_\mu}^2) \right. \nonumber \\
    &\quad \left. - 48 f^2_c \, _2F_1\left(-\frac{13}{6}, -\frac{1}{2}; -\frac{7}{6}; \frac{f^2_c}{{f_\mu}^2}\right)\right] \nonumber \\
    &-\frac{3}{494f^{16/3}} \sqrt{\frac{1 - \frac{{f_\mu}^2}{f^2}}{1 - \frac{f^2}{{f_\mu}^2}}} \times \left[13 \sqrt{1 - \frac{f^2}{{f_\mu}^2}} (f^2 - {f_\mu}^2) \right. \nonumber \\
    &\quad \left. - 48 f^2 \, _2F_1\left(-\frac{13}{6}, -\frac{1}{2}; -\frac{7}{6}; \frac{f^2}{{f_\mu}^2}\right)\right]\,, \\
    \phi_{\rm{MF}}^{s=1}(f) &= \frac{3}{320f^{13/3}_c} \sqrt{\frac{1 - \frac{{f_\mu}^2}{f^2_c}}{1 - \frac{f^2_c}{{f_\mu}^2}}} \times \left[10 \sqrt{1 - \frac{f^2_c}{{f_\mu}^2}} (f^2_c - {f_\mu}^2) \right. \nonumber \\
    &\quad \left. - 39 f^2_c \, _2F_1\left(-\frac{5}{3}, -\frac{1}{2}; -\frac{2}{3}; \frac{f^2_c}{{f_\mu}^2}\right)\right] \nonumber \\
    &-\frac{3}{320f^{16/3}} \sqrt{\frac{1 - \frac{{f_\mu}^2}{f^2}}{1 - \frac{f^2}{{f_\mu}^2}}} \times \left[ 10 \sqrt{1 - \frac{f^2}{{f_\mu}^2}} (f^2 - {f_\mu}^2) \right. \nonumber \\
    &\quad \left. - 39 f^2 \, _2F_1\left(-\frac{5}{3}, -\frac{1}{2}; -\frac{2}{3}; \frac{f^2}{{f_\mu}^2}\right)\right]\,.
\end{align}

The total correction on the GW phase is
\be
\Psi_{\rm MF} =2\pi ft_{\rm{MF}}(f)-\phi_{\rm{MF}}(f) \,.
\ee

\bibliography{mbGW}

\end{document}